# Biological Engineering – What does it mean? Where does it (need to) go?

AUTHORS

Ulrike A. Nuber[1,*], Viktor Stein[2,*]

AFFILIATIONS

[1]ulrike.nuber@tu-darmstadt.de, Centre for Synthetic Biology, Dept. of Biology and Dept. of Mechanical Engineering, Technical University of Darmstadt, Schnittspahnstrasse 13, 64287 Darmstadt, Germany

[2]viktor.stein@tu-darmstadt.de, Centre for Synthetic Biology, Dept. of Biology, Technical University of Darmstadt, Schnittspahnstrasse 12, 64287 Darmstadt, Germany

* Correspondence to: Ulrike A. Nuber and Viktor Stein

ABSTRACT

Biological engineering, the convergence between engineering and biology, is at the forefront of significant advances in healthcare, agriculture, and environmental sustainability, making it highly relevant to current scientific and societal challenges. We take a comprehensive look at this broad and interdisciplinary domain, structure it into three main areas – bioinspired, biological and biohybrid approaches – and dissect inherent and fundamental challenges along with opportunities, highlighting specific examples. We describe how data-driven discovery and design, in conjunction with artificial intelligence, can mitigate the absence of reductionist models in these areas. Additionally, we address the education of a new generation of biological engineers, emphasizing mathematical, technical, and artificial intelligence frameworks.

## INTRODUCTION

A capacity for engineering has continuously shaped human civilization, fundamentally changing the way we live, interact, and make use of our environment. Engineering builds on a deep scientific understanding of physical, chemical, and biological phenomena that are applied in a targeted fashion to develop new technologies for the benefit of mankind. Prime examples include numerous highly transformative industries and technologies in the 20th century ranging from civil, mechanical, electrical, and computational engineering to synthetic chemistry and chemical engineering. These were based on early discoveries in physics in the 17th and 18th century in the fields of mechanics and optics, and the rapid advances made in physics and chemistry in the 19th century, including electric and magnetic phenomena, thermodynamics, and atomic theory. In significant parts, these developments have been enabled by developing detailed mechanistic and quantitative scientific frameworks based on well-defined and well-understood physical and chemical principles.

In contrast, a capacity to engineer existing biological systems, hybrid ones consisting of biological and artificial components, or systems inspired by biology, is emerging much more slowly. For once, molecular components and mechanisms underlying fundamental biological processes were only unraveled over the past 50 years. The application of X-ray crystallography and the discovery of the double helix structure of DNA in the 1950s were decisive milestones that laid the foundation for modern molecular biology. Further, and more significantly, biological systems are inherently complex which has frequently eluded reductionist models and poses a significant challenge towards mechanistic modeling and predictive engineering (**Figure 1**). Finally, some key features underlying the potency of biological systems – such as a capacity for self-organization and evolution – elude standardization which is central to the scalability and innovation capacity of many physics- and chemistry-based engineering disciplines. Therefore, biological engineering necessitates a different approach and mindset.

Yet, a broad discussion of what biological engineering stands for is lacking which renders it difficult for a global readership in science, business, and politics to gain a structured overview in this rapidly developing domain with its significant societal and environmental impact.

In this article, we broadly discuss three areas of biological engineering along with their opportunities and inherent challenges associated with them. A particular focus is on how limitations of reductionist models can be mitigated by data-driven designs in association with artificial intelligence. This is contrasting the first-principle-approach underlying chemistry- and physics-based engineering. Finally, we argue for new types of engineers that need to be trained and equipped with a combination of profound technical and mathematical skills and a fundamental understanding of how biological systems function. This also requires experimental methods to study and engineer them, which at the moment, is rarely covered in a comprehensive and integrated fashion in university curricula, but critical to unravel the full potential of biological engineering.

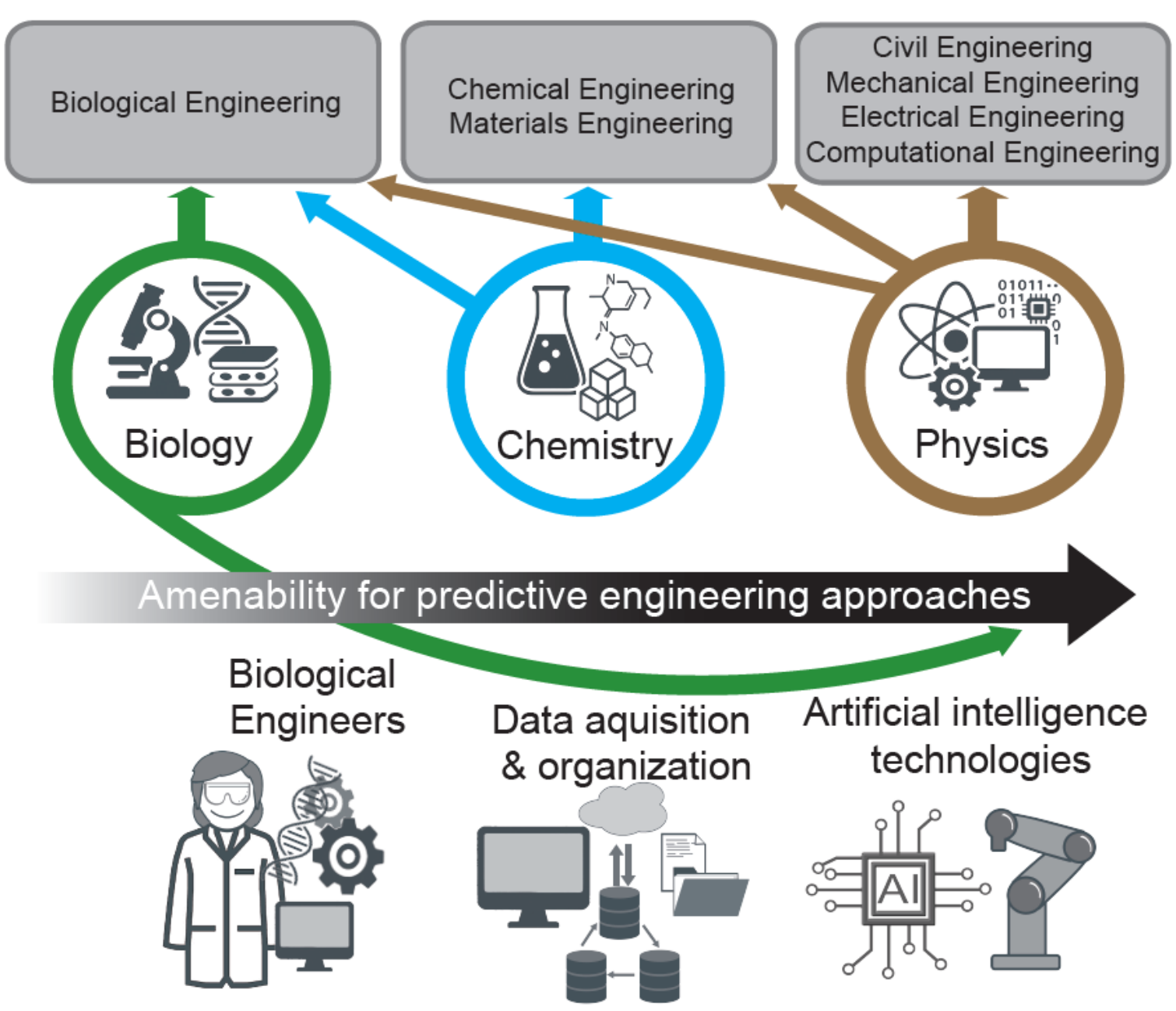

**Figure 1.** Engineering domains and their natural science foundations, physics, chemistry, and biology, with different amenability for predictive engineering approaches. These approaches allow to predict how a system or device will behave before it is physically built or deployed. The extent of predictive capability may vary depending on the type of system. Training the next-generation of biological engineers, standardizing data acquisition and organization, and artificial intelligence technologies will contribute to increasing biology´s amenability for predictive engineering.

MAIN TEXT

**What is Biological Engineering? Different approaches and their potential**

The world is currently facing many challenges, for instance, in relation to the climate and energy crisis, the scarcity of resources along with global health challenges ranging from infectious diseases to ageing societies. However, despite its promise and potential to address some of the most imminent problems, the capacity to engineer desired biological and biology-related or -derived systems still eludes us.[1 2]

Strikingly, living organisms and ecosystems are products of billions of years of evolution, during which they have developed remarkable resilience using principles based on self-organization, regeneration and adaptation. Further, they can readily interface and interact with their environment, which includes living and non-living entities. Most crucially, they can do this autonomously across molecular, micro- and macroscopic as well as vastly different temporal scales. Finally, fundamental features of living entities are their capacity to regenerate and to replicate themselves. This provides models, mechanisms and blueprints for new technologies that one can emulate, build, and interface with, before leveraging them in key areas of need, for instance, to develop smart therapeutics that adapt to dynamically changing disease processes and sustainable production processes that tap into readily available and recyclable resources. Conversely, key properties such as evolution, adaptation, self-organization, regeneration, and replication, that are intrinsic to biology and very much underline its potency, are associated with and based on error and uncertainty which ultimately contrasts and contradicts the robustness and predictability associated with physics or even chemistry engineering strategies.

In order to exemplify its potential but also delineate challenges, we discuss biological engineering in a holistic fashion across three distinct areas – namely, **1.** bioinspired (**Table 1**), **2.** biological (**Table 2**) and **3.** biohybrid (**Figure 2**). We suggest using the term "biological and thereof derived or related systems" when not only referring to purely biological systems that are tailor-engineered, but also to biologically inspired and biohybrid systems that both critically depend on a profound understanding of biology. We note the scope of biological engineering is broader than what at times is considered and discussed in the context of engineering biology or synthetic biology. Notably, it extends beyond their definition proposed at the turn of the 21st century which focuses on applying engineering principles to engineer biological systems. [3] [4] This broader scope of biological engineering is deliberate and aims to illustrate the wider potential of this highly interdisciplinary domain along with the complexities and necessities for any underlying educational frameworks to be firmly grounded in both biology and engineering.

**1. How does Biology inspire Engineering?**

Historically, biology – along with its natural evolution-based innovation algorithm – has served as a great source of inspiration for engineers driving the development of highly sophisticated and robust technical systems across nano- to micro- and macroscopic scales. These can be broadly considered as bioinspired engineering approaches and are based on fundamental biological concepts that are newly implemented with technical components and abiotic as well as biotic materials.

A prominent historical example constitutes the study of bird flight and its application to human flight. Early critical observations, such as the importance of a curved wing section for generating lift, were already documented by Leonardo da Vinci in his Codex on the Flight of Birds (1505). Building on the aerodynamic principles, the Wright Brothers subsequently realized the first controlled, sustained, and powered flight using a fixed wing aircraft in 1903.[5] Strikingly, once aerodynamic principles were understood, the ensuing technical evolution of fixed-wing aircraft based on mechanical systems and man-made, abiotic materials yielded technical implementations that were substantially more potent and more powerful than nature could ever envisage, especially in terms of speed and size. Yet, compared to animal flight, fixed-wing aircraft still have limited hovering and maneuvering abilities.[6] Developing artificial flight modes further, small flying machines with flapping wings that biomechanically mimic animal flight [7] [8] now offer superior performance in terms of precision and energy efficiency.

The ability to miniaturize technical systems, combined with detailed molecular mechanistic insights into biological functions, has made it feasible to emulate biological functions with technical systems at the micrometer or even nanometer scale. Modern examples include artificial nano- and microswimmers that are inspired by the remarkable swimming efficiency of motile natural cells mimicking their structure and propulsion, as well as artificial molecular machines mimicking the principles of movement and energy conversion seen in natural mechanisms.[9] [10] [11]

Another example related to movement is the inspiration provided by the coordinated beating of cilia on the surface of cells in embryonic frog tissues. The normal function of these cilia is to transport mucus and other substances from the surface of the embryo. However, when a small piece of tissue is cut out of frog embryos and placed in an aqueous solution, it organizes itself into a spherical mass of tissue, with the cilia guiding movements of these so-called xenobots.[12]

Finally, bioinspired approaches advance artificial intelligence (AI) technologies in the form of neuromorphic computing – technologies emulating the structure and functioning of the human brain. [13] Of note, introducing key features of the actual processes of natural nerve cells into artificial neural networks – rather than focusing solely on neuron cell bodies (nodes) as traditional approaches do – lowers their computational cost of solving a problem, thereby increasing their efficiency.[14] [15]

**Table 1.** Selection of biologically-inspired engineering approaches without and with AI support. The former range from empirical approaches to mechanistic mathematics and physics models.

| **Biological Phenomenon** | **Engineered System / Device** | **Ref.** |
|---|---|---|
| *Empirical to mechanistic Mathematics and Physics model-based Approaches* | | |
| Bird flight: Curved wing | Fixed wing aircraft | 5 |
| Animal flight: Flapping wings | Flapping wing drones | 6-8 |
| Motility modes: Natural cells | Nano- and microswimmers, Xenobots | 9-12 |
| Neuronal network: Information processing | Neuromorphic computing | 13-15 |
| *AI-Powered Approach* | | |
| Biological material properties: | Bioinspired materials: | |
| • Nacre: Uniquely tough, strong, and lightweight | • Nacre-like structures | 20 |
| • Spider silk: Exceptionally tensile strength and elasticity | • Synthetic proteins forming structures surpassing spider silk properties | 22 |
| Chemotaxis by cells and multicellular organisms to navigate and search in a targeted way | Synthetic systems endowed with intelligence akin to natural target search and navigation strategies | 23, 24 |
| Motility modes: Natural cells | Xenobots with optimized motion and replication | 25 |

As exemplified above, biologically inspired approaches have led to significant technological breakthroughs. However, such approaches still face limitations.[16] On the one hand, these relate to a limited feasibility of some biological aspects for technical solutions, and on the other hand to an insufficient understanding of how the biological functions and characteristics that one draws inspiration from are actually realized by nature.

Concerning the feasibility of biological blueprints for engineering solutions, some unique properties and structures of natural systems are difficult to reproduce technically, especially in terms of the components and material properties.[17] Consequently, the performance levels of biological systems frequently cannot be matched by technical solutions. For instance, regarding resource use and energy consumption, self-healing and self-cleaning mechanisms found in nature (e.g. gecko feet that maintain their stickiness on dirty surfaces). Further, the fabrication and scalability of bioinspired solutions for hierarchical, composite, and multiscale materials are challenging and costly, hindering the replication of sophisticated biological architectures and functions at larger, industrial scales.

Further, our insufficient understanding of biological functions and characteristics poses substantial limitations for translating natural phenomena into bioinspired engineered solutions. In particular, how one can deal with the gaps and limits in our biological knowledge that are critically linked to the inherent complexity and uncertainty of biological systems and thus do not rely on superficial biological analogies in the design of bioinspired systems? The design success will be enhanced by deducing functionalities across a broad range of biological contexts in a comparative fashion rather than relying on isolated biological case studies as done in many bioinspired approaches so far. Further, one can strive to formally describe biological systems in a standardized fashion and thus leverage artificial intelligence and computational modeling to overcome knowledge limitations.

Namely, AI methods promise to accelerate the systematic analysis of biological data and facilitate the abstraction of biological principles. In particular, they can deal with the lack of obvious causal relationships, for instance, based on elementary mechanisms, and with uncertainties. For example, generative adversial networks, as advanced AI models, can replicate the microstructure of elk antlers for which only limited mathematical representations exist.[18] Further, AI methods can optimize the design and manufacturing processes, such as additive ones, for bioinspired technologies. This becomes increasingly apparent in bioinspired generative design.[19] For example, the AI reinforcement learning approach can support the design of structures inspired by the unique tough, strong, and lightweight properties of nacre [20], which is difficult to capture with reductionist models. In terms of the abiotic materials used in bioinspired approaches, AI models can optimize and also generate entirely new properties, structures, and topologies.[21] AI methods can decode complex protein structure-function relationships that often cannot be fully described by reductionist models (see also section 2. below), and predict and design proteins inspired by exceptional properties of naturally occurring ones. For example, inspired by the tensile strength and elasticity of spider silk, synthetic proteins can be designed that exceed the natural model in mechanical robustness. ([22] and references therein). AI methods also contribute to endowing abiotic systems such as robots and smart active microparticles with natural intelligence strategies to respond to environmental stimuli.[23] [24] Finally, combining an evolutionary optimization–based AI method with physics simulation can determine efficient xenobot shapes for movement or specific tasks.[25] Of note, this approach brought forth an unexpected emergent behavior of Xenobots, kinematic self-replication, as side effect of optimizing for other features.

Taken together, the application of AI methods and their combination with mechanistic physics-based models can lead to bioinspired solutions that surpass their natural blueprints, as they are not limited by biological development or natural evolutionary history.

## 2. How does Engineering inspire and shape Biology?

Conversely, it has been a long-sought goal to engineer biological systems in a targeted fashion. Drawing inspiration from rapid advances in physics and chemistry, and thereof based engineering approaches at the turn of the 20th century, French physicist Stephan Leduc first coined the idea of synthetic biology hypothesizing to engineer biological systems based on first principles.[26] With the majority of components and mechanisms yet to be discovered, this was highly visionary but premature for a practical realization. It was not until the turn of the 21st century that the term synthetic biology was reintroduced[3] [4] and a standardized framework for engineering biological systems formulated.[27]

Arguably, one central trademark of modern synthetic biology constitutes the integration of engineering concepts[28] – associated with both engineering processes and technical designs – into the traditional discovery-oriented life sciences that were heavily reliant on exploratory trial-and-error. This shift also attracted many non-biologists such as computational and electrical engineers with their mathematical and technical skill sets to the field.

**Table 2.** Selection of engineering-inspired synthetic biology approaches without and with AI support.

| **Engineering Concept / Infrastructure/ Process** | **Biological System / Device / Tool** | **Ref.** |
|---|---|---|
| *Empirical and First-Principle Designs* | | |
| Abstraction / Standardization | Part, Device, System, Chassis | 27 |
| Microchip Factories / Manufacturing Plants | Biofoundries (for developing “DNA programs” that execute bespoke biological functions)[#] | 29 |
| Computer Aided Design (CAD) Tools | COBRA, BioCAD Tools[#] | 30, 35 |
| Electronic Design Automation (EDA) | Genetic Circuit Design Automation | 32 |
| Electronic Engineering / Control Theory<br>• Key Concept: Integrated feedback for robust and reliable operation under variable conditions | Genetic Circuits for Biomolecular Computation<br>• Specific Example: Antithetic feedback for perfect robust adaptation under cellular noise | 33, 34 |
| *AI-Powered Data Driven Designs* | | |
| - | AI Protein Design<br>• RFDiffusion, ProteinMPNN | 45 |

[#] Methodologically, Biofoundries, COBRA, and BioCAD Tools also carry great future potential for being powered by AI.

Key concepts and processes that have inspired synthetic biology in the construction of complex biological systems especially concern abstraction and standardization as well as iterative design and design automation.

Firstly, drawing inspiration from electronic circuit design, the identity of biological components was abstracted.[27] In this way, a higher order language was created that is necessary and sufficient to design biological systems with minor technical skill and knowledge of the underlying mechanisms and experimental approaches. Accordingly, nucleic acids, proteins and thereof based functional elements such as gene regulatory elements, metabolic and signaling pathways and cellular hosts were hierarchically classified as parts, devices, systems, and chassis. This drive for abstraction also implied that components were modular, compatible, and interoperable – i.e. they were arbitrarily composable and predictively functioned in different type of contexts.

Secondly, a formal description of the biotechnological development cycle – known as the design-build-test and learn (DBTL) cycle – was popularized that effectively broke down the engineering process into standardizable sub-steps while emphasizing the need for design iterations to achieve optimal solutions in the development of bespoke biological systems. This requires **(i)** formulating a design hypothesis, **(ii)** encoding the design into a DNA sequence, and **(iii)** functionally testing the resultant "DNA program" towards the design hypothesis. In practice, depending on the quality of the design hypothesis, this may encompass testing – i.e. effectively screening – a variable number of variants and DBTL iterations to optimize any given "DNA program".

Thirdly, analogous to microchip factories and manufacturing plants, biofoundries are being established for the automated design of bespoke biological systems. These are effectively tasked to develop "DNA programs" that code for a biological function in a particular cell and specific application.[29] This also entailed the development of dedicated software resources and computer aided design (CAD) programs[30] along with the implementation of highly specialized robotic facilities that physically assemble "DNA programs" and subsequently test their function in a standardized and automated fashion.

Further, synthetic biologists have continuously drawn inspiration from distinct theoretical concepts and technical designs. One prominent example includes the profound impact of electrical engineering and control theory[31] on the construction of genetic circuits for (bio)molecular computation. These are typically based on a defined set of well characterised biomolecular components that are assembled into defined networks capable of complex information processing tasks. In one seminal example, the automated design of genetic circuits capable of computing complex logic functions from a library of biomolecular regulators was successfully demonstrated effectively emulating electronic design automation.[32] To this end, the design pipeline first entails a high-level definition of a complex logic function spelt out in Verilog which is then encoded into DNA using a dedicated software tool (termed Cello) and eventually assembled into a "DNA program" using an automated biofoundry.

In another seminal study, a new type of integral feedback – termed antithetic feedback – was developed capable of robust perfect adaptation in the presence of stochastic cellular noise.[33] Remarkably, a biomolecular implementation of antithetic feedback – which has so far not been observed as a regulatory motif in nature – was successfully tested under different experimental conditions and perturbations.[34]

In complementary efforts, mastering complexity, synthetic biologists have drawn on systems level engineering to effectively navigate the complex physiological landscapes of cells. One prominent example constitutes physics-based COBRA (COnstraint-Based Reconstruction and Analysis) framework used in metabolic engineering.[35] These describe complex genome-scale metabolic maps that embed physicochemical laws as hard constraints on reconstructed biochemical reaction networks featuring thousands of metabolites, reactions and enzymes. Notably, COBRA aims to confer a capacity for the rational design of metabolic networks, for instance, predicting genetic modifications that maximize production of a target metabolite and estimating theoretical yields along with trade-offs between growth and product formation. Models are frequently integrated and updated with complex omics data sets to improve predictive power and thus reduce experimental screening efforts.

Despite its continuous and successful pursuit of rendering biology more engineerable, the *a priori* design of biological systems remains a highly challenging endeavor. Arguably, this is primarily due to the complexity of biological systems. In particular, this concerns a lack of quantitative and ultimately reductionist models of biological concepts and phenomena based on primary equations that are frequently key to rational design in chemistry- and physics-based engineering disciplines.

Arguably, the strong context-dependency of biological functions across different scales and levels constitutes a key challenge towards *a priori* design:

Firstly, genetic elements are typically subject to epistatic interactions, which limits their abstraction, standardization and ability to reliably compose them into higher order functional modules. For instance, even for comparatively simple and well understood entities such as a transcription unit, a ribosome binding site and the ensuing open reading frame coding for any given protein along with its codon composition cannot be considered independent entities, but will mutually affect each other through secondary structures and thus result in variable expression levels that are challenging to predict quantitatively, and thus frequently render *a priori* modeling a fundamental challenge.[36]

Secondly, biophysical parameters used to model primary equations can significantly differ in idealized *in vitro* conditions compared to in the crowded *in vivo* environments.[37] [38] Therefore, biological mechanisms and thereof based models frequently remain limited to cartoon drawings that typically capture topology but lack quantitative information that are key to *a priori* modeling. Further, the extent of which any given component can be effectively made, significantly limits the transferability of individual modules across different types of cells and organisms.

Thirdly, any module that is being integrated and interfaced into any type of cell will depend on endogenous cellular processes and resources. These can range from elementary housekeeping functions, such as the translational machinery needed to make proteins, to energy states that will differ across cell types, states, and experimental conditions.[39] In similar considerations, the functional insulation of individual modules constitutes an exquisite challenge as any given integrated module effectively exerts a load (i.e. retroactivity) on upstream processes that will impact cellular physiology and circuit performance itself.[40]

Any of these aspects poses a significant challenge towards the development of reductionist models as the function of any part, device or entire module cannot be considered in isolation but will to a significant degree depend on biological context.

Consequently, the majority of engineering endeavors in synthetic biology still depend on highly empirical development cycles that, for instance, heavily rely on copy and paste from nature in association with screening and/or directed evolution.[41] [42] [43] While powerful, the extent to which any given (bio)molecular component, pathway or cell can be re-engineered in this way, is ultimately limited especially the further any given starting point turns out to be away from the envisaged target function. Further, interoperability of components is frequently limited in non-native operating environments and typically necessitates empiric optimization.

Therefore, a different type of engineering approach is required to support the design of biological systems. Towards this goal, machine learning and AI constitutes the most promising tool (**Figure 1**). Crucially, AI can augment many of the empiric approaches available for engineering biological systems as well as guide the development of new ones. In particular, models can now be developed based on the agnostic analysis of data patterns and correlations independent of mechanistic models that are based on fundamental first principles, i.e. independent of causal relationships based on elementary mechanisms.

Recent, rapid and revolutionary advances in AI-based protein design speak testimony of the potential of AI in biological engineering.[44] [45] Notably, AI-based protein design strikingly demonstrates how a fundamental biological problem that was initially attempted to be solved based on first principles was rapidly outpaced through agnostic analysis of data patterns using AI.[45] [46] This development did however – and this is key – heavily rely on high quality protein structural data sets generated by the scientific community for decades with an estimated net worth of 23 billion USD in 2024 that was predominantly financed by public funds.[47]

While a capacity to design proteins constitutes a highly remarkable achievement in itself, this only provides one piece towards the design of bespoke biological systems. Ensuing is the assembly of biomolecular components into pathways, circuits and networks, and their integration into cell types with distinct physiologies. Owing to the inherent complexity and idiosyncrasy of biological systems, this remains a highly empiric process. Thus, for AI to realise its potential across all scales of biological engineering, new design frameworks need to be developed: Firstly, this requires standardized, experimental analytical frameworks for acquiring sufficiently large, high-quality, and cost-effective datasets that, secondly, are amenable to the development of AI-based models that enable predictions. In terms of experimental analytical frameworks, biology provides one unique feature over chemistry and physics. Namely, for as long as one can resolve and relate any given biological function to its coding DNA program – analogous to maintaining a link between genotype and phenotype underlying natural evolution – one can take advantage of the diminishingly low costs of next-generation DNA sequencing technologies to read-out and thus map the function of biological programs. The key difference to established directed evolution endeavors is that any given engineering process does not have to navigate evolutionary trajectories. Instead, AI combined with a capacity for standardized experimental frameworks promises to build models facilitating design predictions either solely based on the analysis of empiric data[48] or in combination with mechanistic models.[49]

Developing a capacity for the *a priori* design of biological systems is by no means an academic exercise but carries great potential across a range of applications especially as it promises to shorten costly development cycles to address unmet needs using tailored biological systems. For example, in the medical field, the engineering of synthetic signaling networks in human cells enables them to sense and respond to specific factors released by other cells at a fast time scale that could be beneficial for a range of therapeutic and diagnostic applications.[50] Further, bespoke microbial cell factories promise to convert readily available carbon sources into value-added chemicals.[51] Finally, engineering genetic circuits in plants can reprogram plant growth and enhance crop resilience to extreme environmental conditions.[52]

Yet, despite their undisputed potential, any engineered biological system ultimately remains limited to operate within the constraints of biology – i.e. cells mutate upon replication and are in turn subject to evolutionary forces. For instance, cells used as therapeutics accumulate a large number of mutations upon limited *ex vivo* propagation, which potentially contradicts regulatory guidelines requiring well-defined therapeutic agents. Similarly, any environmental release of genetically modified organisms is associated with a containment risk, an issue that is particularly discussed for microorganisms to be released in open environment for bioremediation, biosequestration, pollutant monitoring, and resource recovery purposes.[53] Addressing these challenges with the constraints of biological systems will constitute a significant technological challenge in the future. However, the lack of durability, safety and controllability that can occur with biological systems are features typically offered by synthetic materials and technical systems.

## 3. How can we interface Biology with (abiotic) materials and technical systems - the best of both worlds?

Addressing these limitations, hybrid systems that combine biomolecules, cells or tissues with non-living artificial materials carry great potential. In particular, they promise to combine key advantages of each type of system while mitigating their shortcomings. This encompasses the selectivity and specificity of biological systems—together with their autonomy, adaptability, and regenerative capacity—combined with the mechanical stability, precision, robustness, durability, and long-range communication capabilities of abiotic materials and technical devices. In this way, biohybrid systems with entirely new properties and functions are envisaged, paving the way for highly innovative and even unexpected applications, e.g. the use of biohybrid robots for the autonomous, real-time monitoring of aquatic, terrestrial, and potentially even extraterrestrial ecosystems. Since such biohybrid systems and devices are expected to possess properties that become increasingly similar to the above-mentioned key biological ones (evolution, adaptation, self-organization, regeneration, and replication), definitions and terms like "machines" might need to be revised.[54]

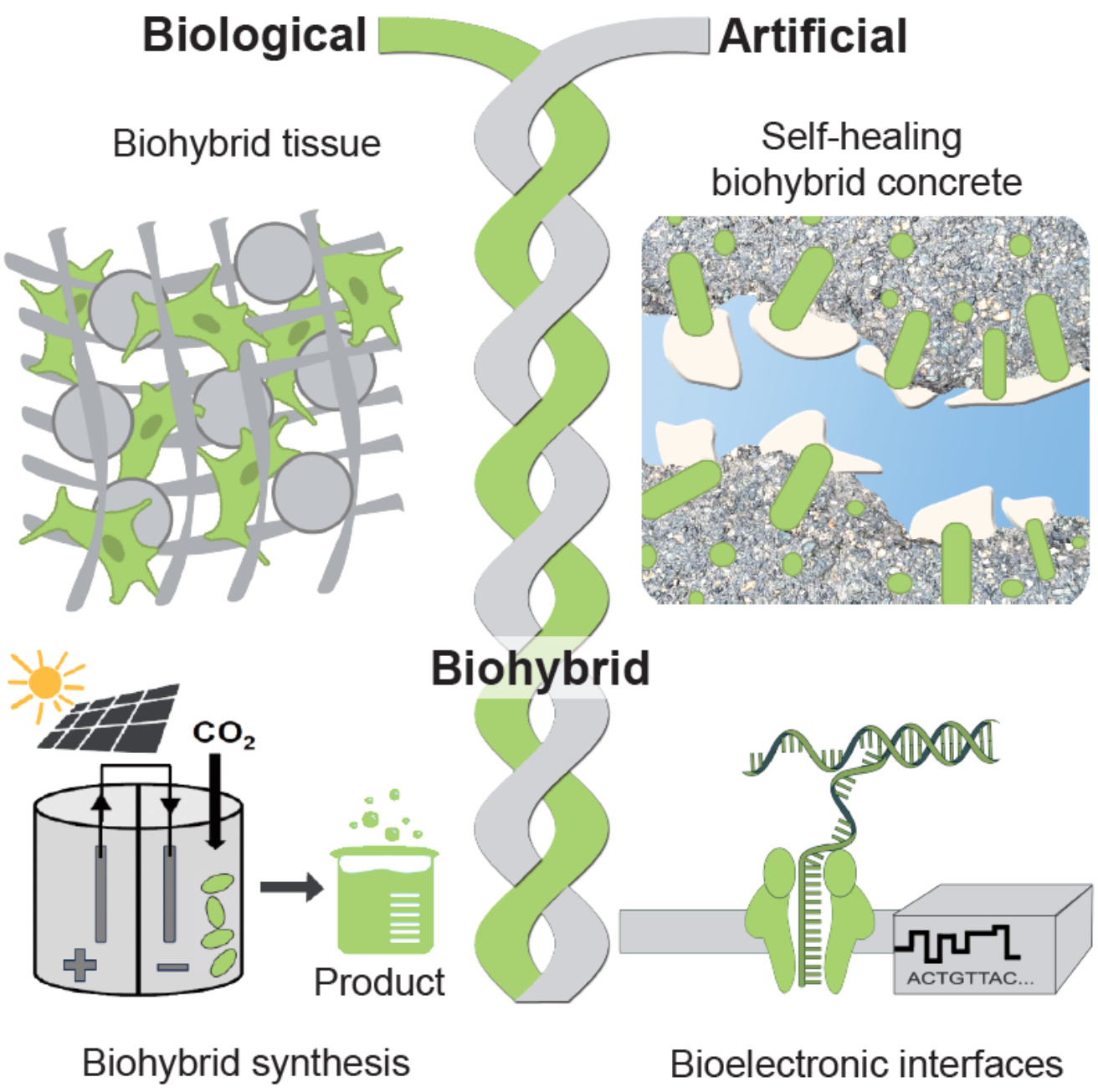

**Figure 2.** Examples of biohybrid systems: Biohybrid human tissue consisting of natural cells, extracellular materials, and artificial cells; Water-filled crack in concrete containing bacterial spores which upon activation produce calcium carbonate and heal the defect; Net-zero (bio)synthesis of matter and materials: A bio-electrochemical interface fueled by solar light facilitates the enzyme-catalyzed synthesis of (bio)chemical products by genetically engineered bacteria; Protein nanopore-based device to sequence DNA.

Biohybrid approaches have long been applied in tissue engineering and have resulted in products used in clinical settings. These approaches blend synthetic materials with cells for implantation or extracorporeal use, or only introduce materials into the human body. A seminal example is artificial skin for closure of burn injury wounds that contains an outer silicone and bottom cell-containing layer.[55] Biohybrid systems designed to replace entire organs such as the liver and kidney are emerging, but not yet in standard clinical use. [56] [57] The integration of materials in the form of electronics in human tissues to monitor, modulate, and enhance tissue functionality[58] overlaps with long existing efforts in another field - bioelectronic interfaces.

In bioelectronic interfaces, the goal is to effectively connect molecular and cellular components with technical devices. Prime examples are neuroelectronic devices that can directly engage in two-way, closed-loop communication with the central nervous system such as the brain or the spinal cord with unprecedented application potential, for instance, treating neurological disorders by compensating for the loss, disturbance or interruption of natural neurotransmission.[59] [60]

Beyond actuation, bioelectronic interfaces also carry great potential in detection, as exemplified by numerous bioelectronic sensors, notably, glucometers that integrate glucose processing proteins to measure glucose levels[61] and protein nanopore-based devices to sequence DNA,[62] and more recently to analyze proteins.[63] These devices take advantage of the exquisite specificity and selectivity of biological receptors along with the sensitivity, programmability, reliability, and the size of micro- or even nano-engineered electronic systems. The resultant devices constitute powerful tools that are frequently capable of differentiating between highly related molecules using portable devices, thus realizing unprecedented access to molecular information without dedicated laboratory infrastructure.

Finally, a key driver underlying the innovation of biohybrid systems concerns the development of sustainable processes for the net-zero (bio)synthesis of matter and materials. Analogous to information processing, the bio-electrochemical interface carries great potential, for instance, fueling the enzyme-catalyzed synthesis of (bio)chemical products using physical energy sources, especially light.[64] In particular, bioelectrodes can functionally replace photosynthesis by providing reducing power in the synthesis of value-added products from low energy carbon sources, especially $CO_2$. Such types of biohybrid materials, which form or assemble the material itself, or modulate the functional performance of the material in some manner are also referred to as engineered living materials.[65]

Similarly, on the macroscopic material scale, the self-regenerating capacity of living biological systems can be exploited in the development of self-healing biohybrid concrete which relies on the integration of bacterial spores into the concrete mix.[66] [67] Upon the formation of cracks and subsequent entry of water, bacteria are activated and produce calcium carbonate ($CaCO_3$), which precipitates within the cracks and seals them. Depending on the bacterial strain, this can even be coupled to bacteria capturing $CO_2$ from the environment, directly linking the self-healing process to $CO_2$ sequestration.

Despite their great potential, biohybrid approaches are still facing many challenges:

Firstly, fabricating hybrid systems is complex given that biological components, in particular cells, need to be maintained in special environments that support their survival and function for long periods of time. Further, unlike conventional abiotic materials, biological systems are inherently variable: phenotypic and metabolic cell states, molecular properties and interactions fluctuate and are strongly influenced by external factors such as temperature and nutrient availability. In addition, biological systems need to be fueled with biocompatible energy sources, such as glucose and amino acids, especially in the case of vertebrate cells.

Secondly, there is a need to establish interfaces through which living and non-living components can interact and communicate. This primarily concerns adjusted abiotic materials, and can be illustrated by biohybrid approaches in tissue engineering and in bioelectronics, the potential of which is still far from being fully exploited. For instance, current artificial materials provide structural support, deliver protein, DNA components, or electrical signals to instruct cells unidirectionally, release signaling components upon certain cell triggers or influence cells biomechanically.[68] However, the next generation of artificial materials interacting with human cells is inspired by cellular features and expected to overcome existing challenges. These challenges include responding to exact cellular demand, operating within precise concentration windows of bioactive molecules, and functioning in complex and changing cellular environments. This requires that entirely new material types with superior interfacing come to life, including supramolecular gels and artificial cells, that are equipped with intelligent functions, longevity and long-term biocompatibility.[69] [70] [71] Modeling and designing such interfaces constitutes a major challenge. Emergent behavior of biohybrid microrobots, such as a micrometer-sized artificial particles conjugated with bacteria and a bacterium conjugated with nanoparticles, have been modeled using statistical simulation[72]. However, reductionist and multiphysics models alone can have limitations in fully capturing biohybrid system interfaces because they may miss emergent properties that are not predictable from the sum of living and non-living parts, context-dependence, complex feedback, and multiscale causality[73], all of which define biohybrid system performance and adaptation.

Thirdly, regulatory bodies and pathways must be defined for hybrid entities that are made up from new combinations of living and nonliving components, presenting unique governance, ethical, and legal challenges. This is relevant for both biomedical and environmental applications, especially if genetically modified organisms are potentially released outside of certified laboratories.

Fourthly and finally, for AI technologies to effectively navigate the complex parameter space, to connect knowledge across various fields in the biotic and abiotic world, and to facilitate the underlying development process of biohybrid approaches, new analytical frameworks with unifying definitions and standardized experimental protocols are required for generating data that is finally AI ready. This is a prerequisite for AI methods to overcome limitations of reductionism and providing a framework capable of reflecting the dynamic complexity and emergent properties essential for successful biohybrid system design and deployment. Given the idiosyncrasy and its highly interdisciplinary nature of biohybrid approaches, this constitutes a formidable challenge and will vary for different types of biohybrid systems.

Efforts are well underway, for instance, machine learning tools allow an automated extraction of information from text and an extraction of relationships to assess the biocompatibility of materials used in combination with human tissue based on retrospective experimental studies.[74] Further, the design and fabrication of hydrogels for applications in wound healing, biosensing, drug delivery, and tissue engineering is supported by integrated AI approaches that advance capabilities beyond those offered by reductionist and multiphysics models alone.[75]. Here, machine learning methods analyze large datasets to identify hydrogel components and formulations and their relationships with experimental outcomes, and support the design of entirely new materials.

With no doubt AI technologies applied in biohybrid approaches will ultimately have to evolve in a highly domain-specific fashion to serve the vastly different areas.

## The road ahead – implications for developing individual academic expertise

Given the undisputed potential of biological engineering in all its facets, the question arises how can biological engineering be driven forward? In particular, how can the next generation of biological engineers be trained effectively, and what type of institutional structures are required to support these endeavors in the best possible way?

Future graduates in biological engineering need to be uncompromisingly equipped with an engineer's mindset enabling them to think, study, and develop new technologies and thereof based products in a holistic fashion. Ideally, this covers both **(i)** basic research into fundamental technologies and **(ii)** translational aspects covering the entire innovation cycle from ideation to its practical implementation and fabrication as well as entrepreneurship. This significantly contrasts with the discovery-oriented approach in natural science – focused on analysis rather than synthesis – which aims for a detailed mechanistic understanding of natural phenomena. Further, it is crucial to equip biological engineers with a strong mathematical and computational skill set which is key to develop predictive and especially scalable design frameworks as has been the key to success of classical engineering disciplines. Yet, owing to the complexity of biological systems and the concomitant lack of reductionist models, design frameworks will in significant parts be rooted in artificial intelligence. In order to effectively train them, they require high quality and especially standardized data sets. Consequently, biological engineers must be well-trained in experimental approaches. In this context, a particular emphasis must be on the development of standardized experimental procedures and protocols that are amenable to AI analysis. Profound knowledge of experimental approaches – which is best acquired through an in-depth practical training – will also enable biological engineers to develop a strong intuitive understanding of biological principles which is key to any of the three approaches of biological engineering.

These notions are supported by an impressive in-depth analysis of the training needs for the next generation of biological designers to effectively tackle and engage with relevant tasks and challenges in the greater area of synthetic biology and engineering biology.[76] The analysis comprehensively compared and opposed industry needs with 90 curricula at 60 academic institutions related to biological engineering worldwide. According to this analysis, there is an imminent need to co-develop both quantitative technical skills with dedicated experimental expertises early in the curriculum. Yet, this critical combination of skills and expertise is currently lacking in most university curricula and instead has to be acquired over the course of lengthy doctoral and post-doctoral training periods. This effectively delays entry into the workforce and career progression from an individual perspective and hampers effective engagement in rapidly developing and changing technoeconomic environments. Similarly, this analysis can serve as a blueprint with equivalent goals and measures readily extrapolatable to training graduates in bioinspired and biohybrid approaches. In particular, it is paramount to co-develop quantitative technical skills along with a fundamental understanding of biology.

Complementing new frameworks for graduate education, it is also imperative to establish dedicated university structures that support the development of biological engineering as a distinct engineering discipline. This constitutes an exquisite challenge given that biological sciences are typically organized and run separately from engineering as well as computer sciences departments. Notably, there are a few academic sites where such developments have been implemented with a highly integrative character – namely, the *Department of Bioengineering* at Imperial College London (https://www.imperial.ac.uk/bioengineering) and the *Biological Engineering* Department at MIT (https://be.mit.edu/) which can both serve as blueprints for potential university structures. Other academic institutions that have already established programs and departments in biological engineering will also readily benefit from a further integration of computer science and AI technologies.

For initiatives to succeed in the long term, it is also critical to bring together faculty members with a highly collaborative mindset and openness to drive research at the interface between biology, engineering, and computational science. Ideally, this is supported by dedicated infrastructure to ensure a continuous and close exchange across research groups as well as disciplinary boundaries. This requires laboratory infrastructure for large scale experimental analysis to get AI ready data and biofoundries for synthesis. Further, it is paramount to support research efforts with coherent curricula. In biological engineering, these should feature a strong emphasis on mathematical and computational skills, in particular at the Bachelor´s level, that will be key to develop domain-specific AI expertise. In addition, fundamental concepts and technologies from classical engineering disciplines must be integrated. Finally, any academic biological engineering structures should ideally be embedded in an entrepreneurial ecosystem with infrastructure and a thriving innovation community. A comprehensive survey of a large number of companies on desired skillsets of genetic engineers (applying our "biological approach") revealed a flexibility to work across organisms, mathematics and computation-guided design capabilities, and a strong basic understanding of molecular biology methods, among others.[76]

How can all these requirements be reconciled and, above all, how can the broad spectrum of engineering sciences be integrated with biology, other natural sciences, and AI content in a new curriculum with specified module restrictions in terms of credits and schedule?

A viable solution is for universities to offer distinct specializations in specific areas of biological engineering. This can be achieved in later semesters at master's level after a common basic bachelor's degree program. Arguably, academic environments that already offer a broad range of natural science and engineering fields are primed to initiate new biological engineering departments. However, the breadth and focus in these fields varies at different universities, naturally necessitating a specialization in the specific type of biological engineering. For instance, combinations of medical faculties and electrical engineering afford a focus on health and biomedical applications. Conversely, agriculture and civil engineering afford a teaching and research focus on environmental applications and sustainability.

To conclude, biological engineering carries great potential, but its implementation constitutes a complex interdisciplinary undertaking. The term "engineering" is rooted in the Latin word "ingenium," meaning inventiveness, and the Middle Latin word "ingeniare," meaning to invent, devise, or contrive. The core root "gen-" conveys the idea of "to generate, produce". Taking a holistic and integrative approach to implementing biological engineering will require an inventive spirit and productivity oriented towards solutions. This concerns the convergence of different research fields and the concomitant development of teaching curricula at the undergraduate and graduate levels.

ACKNOWLEDGEMENTS

We thank Michaela Becker-Röck for her dedicated support in creating the graphics, and Gerhard Thiel, Heinz Koeppl, Benno Liebchen, Felicitas Pfeifer, and Simon Poppinga for critically proofreading the manuscript.

AUTHOR DECLARATIONS

**Conflict of interest**

The authors have no conflicts to disclose.

**Author Contributions**

These authors contributed equally: Ulrike A. Nuber and Viktor Stein

Writing/original draft preparation, Visualization: Ulrike A. Nuber and Viktor Stein

**Funding Sources**

U.A.N. kindly acknowledges the financial support by the LOEWE Research Initiatives Network by the German federal state of Hesse (research cluster FLOWFORLIFE, reference number LOEWE/2/14/519/03/07.011(0002)/78).